\documentstyle[12pt]{article}
\begin{document}
\title{The QCD triple Pomeron coupling from string amplitudes}
\author{A.Bialas\thanks{Institute of Physics, Jagellonian
University, Reymonta 4,  30-059 Cracow, Poland}, H.Navelet and R.Peschanski\thanks{CEA, Service de Physique
Theorique, CE-Saclay, F-91191 Gif-sur-Yvette Cedex, France}}
\maketitle
\begin{abstract}
Using the recent solution of the triple Pomeron coupling in the 
QCD dipole picture as a closed string amplitude with six legs, its analytical form in terms of hypergeometric functions and numerical value are derived.
\end{abstract}

{\bf 1.} The triple Pomeron coupling has since a long time catalyzed a series
of phenomenological and theoretical problems. In practice, it is associated to high-mass diffractive scattering at high-energy which has been observed experimentally and studied in various hadron-hadron reactions. More recently, the observation of high-mass diffraction at the virtual photon vertex of lepton-proton interactions at HERA \cite {he1} has led to a new domain of investigation where the non-negligeable virtuality $Q^2$ of the photon implies a (at least partial) connection with perturbative QCD. On the theoretical point of view, the coupling between three Pomerons in an S-Matrix framework has its counterpart in the six-gluon amplitude in QCD which, in the framework of the BFKL dynamics \cite {li1} has been derived in Refs. \cite {ba1} and obtained in Ref. \cite {pe1} as a  $1\!\rightarrow \!2$ dipole vertex in the related ($1/N_c$ limit of the) QCD dipole framework \cite {mu1}. Recently we reported \cite {bi1} a derivation of the high-mass diffractive dissociation which shows explicitly the implication of the QCD triple Pomeron coupling in the evaluation of the physical  diffractive cross-section. The result could be interpreted \cite {bi1} as  follows:
\begin{equation}
G_{3\cal P}^{eff}(k) \approx \frac 1k \ g_{3\cal P}\left(\frac {\alpha}{\pi}\right)^3 \frac {N_c}4 \left(\frac {2a(y)}{\pi}\right)^3,
\label {2}
\end{equation}
where $k$ is the momentum transfer at the proton vertex, $G_{3\cal P}^{eff}$ represents the effective triple-Pomeron coupling in an S-matrix phenomenological framework. The factor 
$\left(\frac {2a(y)}{\pi}\right)^3 = \left\{7/2\ \alpha N_c \zeta (3) y\right\}^{-3}$ is a logarithmic correction to the Pomeron behaviour $e^{\alpha_P y}$ as a function of the rapidity $y$ and the QCD triple Pomeron coupling  $g_{3\cal P}$  is given by $I_{3\cal P}(1,1,1),$ where \cite {na1}
\begin{eqnarray}
I_{3\cal P}(h_1,h_2,h_3)  =
 \int {d^2\rho_1}{d^2\rho_2}{d^2\rho_3}\ 
{\mid\rho_1\mid^{-h_1+h_2+h_3-2} \mid
1-\rho_1\mid^{-h_3}} \times \nonumber \\ 
{\mid\rho_2\mid^{h_2-2} \mid
1-\rho_2\mid^{h_2-2}}  
{\mid\rho_3\mid^{-h_1} \mid
1-\rho_3\mid^{-h_1}} \mid
1-\rho_1\rho_2\rho_3\mid^{h_1-h_2+h_3-2} ,
\label{4}
\end{eqnarray}
where $h_i = 1 + 2i\nu_i, {i=1,2,3}$ are the complex conformal dimensions of the three $SL(2,{\cal C})$ eigenvectors \cite {li2}  associated to the  three coupled Pomerons
\cite {pe1}. In the same paper \cite {bi1}, another integral $g_{2\cal P}$  multiplies the triple Pomeron coupling  for describing the coupling of  two of the Pomerons to  the target in the particular  case of a dipole target. One finds
\begin{equation}
g_{2\cal P}= \int {d^2\rho_1}{d^2\rho_2}
\ {\mid\rho_1\mid^{-1} \mid
1-\rho_1\mid^{-1}} \nonumber \\ 
{\mid\rho_2\mid^{-1} \mid
1-\rho_2\mid^{-1}} \mid 
1-\rho_1\rho_2\mid^{-1} .
\label{6}
\end{equation}
The purpose of our paper is to give both an analytical formula and a numerical evaluation of the integrals $g_{2\cal P}$ and $g_{3\cal P}.$ As we shall see this requires an extensive use of the recently shown connection \cite {pe1} between QCD dipole vertices and Shapiro-Virasoro amplitudes \cite {sh1} which describe the tree-level amplitudes of a closed string theory.

The plan of the paper is the following: in section {\bf 2}, we recall the connection of the QCD triple Pomeron coupling function with the adequate string amplitudes and give the string interpretation of $g_{2\cal P}(\gamma)$ and $g_{3\cal P}.$ In section {\bf 3}, we use the known relation between closed and open strings to express the quadruple (resp. sextuple) integrals corresponding to $g_{2\cal P}$  (resp. $g_{3\cal P}$ in terms of  products of double (resp. triple) integrals in $[0,1]$ real intervals. This gives an immediate answer for $g_{2\cal P},$
while the more involved solution for $g_{3\cal P}$ is treated in section {\bf 4} using a topological characterization of the open string amplitudes. Conclusions and some mathematical by-products
of the obtained relations involving generalized hypergeometric functions are proposed in the final section {\bf 5} 

{\bf 2.} In the paper \cite{pe1}, a general relation was established between
$1\!\rightarrow \!p$ dipole vertices and dual
Shapiro-Virasoro amplitudes $B_{2p+2}$ describing tree level amplitudes of a  closed string theory. In particular, the $1\!\rightarrow \!2$ vertex, which is equivalent to the triple Pomeron coupling in the $1/N_c$ limit of BFKL dynamics, can be identified with the $B_{6}$ Shapiro-Virasoro \cite {sh1} amplitude with appropriate variables. Indeed, one writes in the Bardacki-Ruegg formalism \cite {br1,fr1}:
\begin{equation}
B_N\ = \int \prod\limits_{i=2}^{N-2}d^2\rho _{i} \prod\limits_{i=2}^{N-2}\mid \rho _{i}\mid^{2K_{1i}}
\prod\limits_{2\le i < j \le N-1}\mid 1-\rho _{i}\cdot\cdot\cdot\rho _{j-1}\mid^
{2K_{ij}},
\label{8}
\end{equation}
where the exponents $2K_{ij}$ play the r\^ole of scalar products of external momenta in the string framework. Note that the definition of  $K_{ij}$ can be extended \cite {fr1} to the full set of indices $\{ i = 1,\cdot\cdot\cdot,N\}$ by the conditions
\begin{equation}
\sum\limits_{i=1}^{N}K_{ij}=0\ ;\ K_{ii}=2,
\label{10}
\end{equation}
which allow one to put the expression for $B_N$ in a completely cyclic  and $SL(2,{\cal C})$-symmetrical form
known as the Koba Nielsen conformal invariant parametrization \cite {ko1,fr1}
\begin{equation}
B_{N}=\int \prod\limits_{i=1}^{N}d^2z _{i}\ \left[ \frac{dz_{\alpha }dz _{\beta }dz _{\delta }}{\left| z _{\alpha \beta }z _{\beta \delta }z _{\delta \alpha }\right| ^{2}}\right]
^{-1}\prod\limits_{i<j}^N
\ \mid z _{ij}\mid^{2K_{ij}} 
,  \label{12}
\end{equation}
where $z _{kl}=z _{k}-z _{l},$ and  the factor
$\left[\cdot\cdot\cdot\right]^{-1}$ corresponds to the formal division by the infinite ``volume'' of the invariance group $SL(2,{\cal C})$ leaving only $N-3$  variables for integration, the other ones $\{z _{\alpha} ,z _{\beta }, z _{\delta }\}$  choosen    arbitrarily. The choice $\{0,1,\infty\}$ and a suitable reparametrization of variables \cite {fr1} ensures the equality of formulae (\ref{8}) and (\ref{12}).

Using definition (\ref{8}) and the expressions (\ref{4}) and (\ref{6}), it is easy to identify the exponents $K_{ij}$ corresponding to the Pomeron couplings. In particular one finds:
\begin{eqnarray}
g_{2\cal P}\!\!&=&\!\!B_5 \left\{K_{ij}= -\frac 12 \  \forall  i \ne j;\ K_{ii}
=2\right\} \nonumber \\
 g_{3\cal P}\!\!&=&\!\!B_6 \left\{K_{12}\!\!=\!\!K_{34}\!\!=\!\!K_{45}\!\!=\!\!0;K_{ij}\!=\!-\!\frac 12 \ \forall (ij) \ne (12), (34), (56);K_{ii}\!
=\!2\right\}
\label{14}
\end{eqnarray}
A comment is in order about the choice of the set $K_{ij}$ in formula (\ref{6}).
It is well-known \cite {fr1} that $B_N$ reflects the {\it sphere} topology  of the surface spanned by the closed string described by the tree-level  amplitude. As such, the couples  $(ij), K_{ij}=0$ in $B_6$ are arbitrary partitions in pairs of the six indices. The choice in formula (\ref{14}) is thus for convenience. As we shall see, this arbitrariness is not true for the {\it disk} topology of the open string amplitudes $A_6$ which we will now consider.

{\bf 3.} There exists a remarkable relation between closed and open string tree-level amplitudes \cite {ka1},
\begin{equation}
B_N = \sum\limits_{{\cal O},{\cal O'}}\ A_N({\cal O})\ A_N({\cal O'})\ e^{i\pi f \left({\cal O},{\cal O'}\right)},
\label{16}
\end{equation}
where $\cal {O},\cal {O'}$ denote the  possible independent orderings of the external legs of an open string tree-level amplitude on the boundary of a disk which represents the world-surface spanned by the open string. Indeed, only non-equivalent orderings are those which are not related by cyclic and/or reversal symmetry \cite {fr1}. The phase factors $e^{i\pi f\left(\cal {O},\cal {O'}\right)}$ are defined \cite {ka1} by the appropriate analytic continuation in the complex plane of the integration variables.  one writes
\begin{equation}
A_N = \int_0^1 \prod\limits_{i=2}^{N-2}d\rho _{i} \prod\limits_{i=2}^{N-2}\left(\rho _{i}\right)^{K_{1i}}
\prod\limits_{2\le i < j \le N-1}\left( 1-\rho _{i}\cdot\cdot\cdot\rho _{j-1}\right)^
{K_{ij}},
\label{18}
\end{equation}
and, for a given ordering  $\cal {O}, A_N(\cal {O})$ is obtained from (\ref{18})
by the permutation $\rho _{i} \Rightarrow \rho _{{\cal O}(i)}.$ It is crucial for our calculation that the amplitudes $B_N$ are thus reduced to products of  $N-3$ dimensional integrals on the segment $[0,1].$ 

Let us now write the results for the amplitudes $B_5$ and $B_6.$ After proper definition of the integral  contours and phase factors, the results for generic amplitudes \cite {ka1} are as follows:
\begin{eqnarray}
B_5 =\!\!&&\sin \pi K_{12} \sin \pi K_{34} A_5 ({\bf 12345})  A_5 ({\bf 21435})\nonumber \\  
     \!\!&& +\sin \pi K_{13} \sin \pi K_{24} A_5 ({\bf 13245}) A_5 ({\bf 31425}),
\label{20}
\end{eqnarray}
and 
\begin{eqnarray}
B_6 =\!\!&&\sin \pi K_{12}\ \sin \pi K_{45}\ A_6 ({\bf 123456}) 
 \nonumber \\
 \!\!&&\times\left \{A_6 ({\bf 215346}) \sin \pi K_{35}
      +A_6 ({\bf 215436})\ \sin \pi \left (K_{13} +  K_{35}\right) \right\}\nonumber \\ 
\!\!&& \ \ \ \ \ +{\rm permutations}\ {\rm of} \ ({\bf 234})\ .
\label{22}
\end{eqnarray}

From (\ref{20}) and the symmetry of the  exponents corresponding to  $g_{2\cal P},$ see (\ref{14}), it is clear that  the amplitudes $A_5(\cal {O})$ are all equal. The solution of the integral~(\ref{6}) can thus be expressed as follows: 
\begin{eqnarray}
g_{2\cal P}&&= \ 2 \ \left\{ A_5 ({\bf 12345})\right\}^2 \nonumber \\
&&= \ 2 \left\{\int_0^1 {d\rho_1}{d\rho_2}
\ \rho_1^{-\frac 12} \left(1-\rho_1\right)^{-\frac 12} \ 
\rho_2^{-\frac 12} \left(1-\rho_2\right)^{-\frac 12} \left(1-\rho_1\rho_2\right)^{-\frac 12}\right\}^2 \nonumber \\
&&=\ 2 \pi \left\{\sum_m \left[\frac {\Gamma(\frac 12 +m)} {\Gamma(1 +m)}\right]^3\right\}^2
=2 {\pi^4} \  \left\{_3F_2(\frac 12,\frac 12,\frac 12;1,1;z=1)\right\}^2 \nonumber \\
&&\equiv \ \frac 1{8\pi^2}\Gamma^8(1/4)\ \approx \ 378.145...,
\label{24}
\end{eqnarray}
where $_3F_2(z)$ is the well-known generalized hypergeometric function.

The case of $g_{3\cal P}$ is less straightforward, since the appropriate conditions (\ref{14}) introduce a partial dissymetry between the exponents.
While the topology of the sphere inherent to the dual amplitude $B_6$ leaves   the zero exponents (e.g. $K_{12}\!\!=\!\!K_{34}\!\!=\!\!K_{45}\!\!=\!\!0$ in formula (\ref{14})) arbitrary,  the various amplitudes $A_6$ appearing in formula  (\ref{22}) and characterized by the disc topology depend on this choice. By a simple investigation of the topologically different cases, see Fig.1, one is led to five  a-priori independent configurations, namely
\begin{eqnarray}
A_6^I &= \ A_6 ({\bf 123456}) \ \ \ A_6^{II} &=  \ A_6 ({\bf 135462})\cr
 A_6^{III} &= \ A_6 ({\bf 124653}) \ \ \ A_6^{IV} &= \ A_6 ({\bf 135246})\cr
                   &      \ \ \ \    A_6^{V} = A_6 ({\bf 146253}). &  
\label{26}
\end{eqnarray}
Indeed, as shown in Fig.1, these are the only unequivalent topological orderings when a particular partition of the set ${\bf 123456}$ into pairs
is singled out with zero exponents, the other being equal and non-zero. 

However, the remaining symmetry of the exponents expressed by the relations  (\ref{14}) is useful to restrict further the number of independent amplitudes 
$A_6$ to be computed. By considering the ``closed-open'' string relations (\ref{22}) starting with the different orderings  (\ref{26}) in the first place (remember that this redefinition is allowed by the sphere topology of the initial $B_6$) one  gets the remarkable relations
\begin{equation}
A_6^{IV} = 2  A_6^{II}\ ;\ A_6^{V} = 2  A_6^{III} = 4  A_6^I\ .
\label{28}
\end{equation}
Finally using these relations to simplify the expression (\ref{22}) for  $g_{3\cal P}$ one can write
\begin{equation}
g_{3\cal P} \equiv \ 4 \ A_6^{II}\ \times \ A_6^{V},
\label{30}
\end{equation}
or any other equivalent expression involving the different  amplitudes related
by (\ref{28}).

{\bf 4.} 
Using the resulting relation (\ref{30}) and the appropriate definitions (\ref{18}), one finds the following expressions
\begin{eqnarray}
A_6^{II}&&= \int_0^1 \prod\limits_{i=2}^{4}d\rho _{i} \prod\limits_{i=2}^{4}\left(\rho _{i}\right)^{-\frac 12}\ \prod\limits_{i=2}^{4}\left(1-\rho _{i}\right)^{-\frac 12}
\left( 1-\rho _{2}\rho _{3}\rho _{4}\right)^
{-\frac 12}  \nonumber \\
A_6^{V}&&= \int_0^1 \prod\limits_{i=2}^{4}d\rho _{i} \ \left(\rho _{2}\right)^{-\frac 12}\left(\rho _{4}\right)^{-\frac 12}
\prod\limits_{i=2}^{4}\left(1-\rho _{i}\right)^{-\frac 12}\ 
\nonumber \\
&&\ \ \ \times \ \left(1-\rho _{2}\rho _{3}\right)^{-\frac 12}\left( 1-\rho _{3}\rho _{4}\right)^
{-\frac 12}, 
\label{32}
\end{eqnarray}
for which both   analytic expressions and numerical evaluations can be performed. 

First, by mere expansion of $\left( 1-\rho _{2}\rho _{3}\rho _{4}\right)^
{-\frac 12}$ one gets:
\begin{equation}
A_6^{II} \equiv \pi^3 {_4F_3 (\frac 12,\frac 12,\frac 12,\frac 12;1,1,1;z=1)}\ ,
\label{34}
\end{equation}
where $_4F_3(z)$ is the well-known generalized hypergeometric function. The evaluation of the amplitude $A_6^{V}$ requires more care due to the coupling factors between all integration variables
. Keeping aside the integration over $\rho _{3}$ in expression (\ref{32}), one writes
\begin{eqnarray}
A_6^{V}&&= \pi^2 \int_0^1 d\rho _{3} 
\left(1-\rho _{3}\right)^{-\frac 12} \left\{{_2F_1 (\frac 12,\frac 12;1;\rho _{3})}\right\}^2\nonumber \\
&&\equiv \ \pi^2 \int_0^1 d\rho _{3} 
\left(1-\rho _{3}\right)^{-1}  {_3F_2 \left(\frac 12,\frac 12,\frac 12
;1,1;\frac {-\rho _{3}^2}{4(1-\rho _{3})}\right) },
\label{36}
\end{eqnarray}
using a known identity between hypergeometric functions \cite {sl1}. The Mellin-Barnes representation yields
\begin{eqnarray}
A_6^{V}&&= \int  \frac {ds}{2i\pi}\frac {\Gamma^2(-s) \Gamma^4(\frac 12 + s)}{\Gamma^2(1 + s)}\ \equiv \ 
G^{2,4}_{4,4} \left(-1\mid \matrix{\frac 12&\frac 12&\frac 12&\frac 12\cr 0&0&0&0\cr}
\right)
\nonumber \\
&&\equiv 4 \sum_m \frac {\Gamma^4(\frac 12 + m)}{\Gamma^4(1 + m)}\left( \psi (1+m) - \psi (\frac 12 + m) \right),
\label{38}
\end{eqnarray}
where $G^{2,4}_{4,4}$ is the Meijer function \cite {pr1} and the $\psi$ functions come from the residues of the doubles poles in the integration over $s.$ Note that $A_6^{V}$ can also be expressed as a derivative of a $_4F_3$ with respect to the parameters
\begin{equation}
A_6^{V}=4\pi^2\ \frac {\partial}{\partial\epsilon} \left\{{\frac {\Gamma\left(\frac12-\epsilon\right)} {\Gamma\left(1-\epsilon\right)} } \ _4F_3(\frac12-\epsilon,\frac12,\frac12,\frac12;1-\epsilon,1,1;1)\right\}_{\epsilon=0}.
\label{39}
\end{equation}

All in all the final result reads
\begin{eqnarray}
g_{3\cal P} &&\equiv 4 A_6^{II} \ A_6^{V}= 16 \pi^5 \left\{\sum_m \frac {\Gamma^4(\frac 12 + m)}{\Gamma^4(1 + m)\Gamma^4(\frac 12)}\right\} \times \nonumber \\
&&\times  \left\{\sum_n \frac {\Gamma^4(\frac 12 + n)}{\Gamma^4(1 + n)\Gamma^4(\frac 12)}\left( \psi (1\!+\!n) - \psi (\frac 12\!+\!n) \right)\right\}\approx 7766,
\label{40}
\end{eqnarray}
where we have explicitely written the expression of the ${_4F_3}$ in (\ref{34}). Equation (\ref{30}) was quoted in our previous paper \cite {bi1}. It coincides with expressions found independently using  a different method in \cite {ko2} and with the numerical evaluation of \cite {bb1}.

{\bf 5.} Some comments of physical and mathematical nature are in order. On the physical ground, it has been shown in \cite {bi1} that the couplings $g_{2\cal P}$ and principally $g_{3\cal P},$ the triple Pomeron coupling in the QCD dipole picture appear in the calculation of the high-mass diffraction cross-section off the virtual photon, for which interesting data taken at HERA exist. It is quite remarkable that the high value obtained for  
$g_{3\cal P}$ may compensate the smallness of the perturbative  six-gluon factor $\left(\frac{\alpha}{\pi}\right)^3$ and give rise to a large effective coupling $G_{3\cal P}^{eff}$ of (\ref{2}). As noticed in \cite {bi1}, this may help in predicting a quite sizeable cross-section, opening the way towards an unified description of diffractive and non-diffractive proton structure functions \cite {ro1}. 

On a more mathematical point of view, the relation between string amplitudes and specific hypergeometric functions show some interesting features.
It has been noticed \cite {na2} that integrals of the type 
\begin{equation}
I_{p+1}(z)\ = \int \prod\limits_{i}d^2u _{i} \mid u _{i}\mid^{\alpha_{i}}
\mid 1-u _{i}\mid^{\beta_{i}}
 \mid 1-\prod\limits_{j}u _{j}z\mid^{\alpha_{0}}
\label{42}
\end{equation}
can be expressed explicitely in terms of known combinations of hypergeometric functions $_{p+1}F_p (z).$ When some coefficients $\alpha$ and $\beta$ are equal
degeneracies may occur which leads to the appearance of derivatives. In this respect, we note that, using (\ref{39}), our result (\ref{40}) may also be expressed as
\begin{eqnarray}
&&g_{3\cal P}(1) \equiv 16 \pi^5\   _4F_3(\frac12,\frac12,\frac12,\frac12;1,1,1;1) \times \nonumber \\
&&\times \frac {\partial}{\partial\epsilon} \left\{{\frac {\Gamma\left(\frac12-\epsilon\right)} {\Gamma\left(1-\epsilon\right)} } \ _4F_3(\frac12-\epsilon,\frac12,\frac12,\frac12;1-\epsilon,1,1;1)\right\}_{\epsilon=0}.
\label{44}
\end{eqnarray}

In order to illustrate how powerful is the method based on string amplitudes, we comment on the mathematical identities implied by the  (\ref{28}) which lead to highly non trivial  relations between special functions. Among them, the relation $A_6^{IV} = 2  A_6^{II}$ leads to the identity
\begin{equation}
\int_0^1 x^{-\frac12}(1-x)^{-\frac12}{_2F_1\left(\frac12,\frac12,1;x\right)}
\ {_2F_1\left(1,\frac12;\frac32;x\right)}\equiv 2\ {_4F_3 (\frac 12,\frac 12,\frac 12,\frac 12;1,1,1;1)}.
\label{46}
\end{equation}
The relation 
$A_6^{V} = 2  A_6^{III}$ gives an integral identity between squares of elliptic functions, namely
\begin{equation}
\int_0^1 (1-x)^{-\frac12}\ \left\{_2F_1\left(\frac12,\frac12,1;x\right)\right\}^2
\equiv 2 \int_0^1 x^{-\frac12}\ \left\{_2F_1\left(\frac12,\frac12,1;x\right)\right\}^2,
\label{48}
\end{equation}
where the functions $_2F_1$ appear after partial integration over $\rho_2$ and $\rho_4$ in formula (\ref{32}), and the elliptic function of first kind ${\bf K}(k) = \frac {\pi}2\ _2F_1^2\left(\frac12,\frac12,1;k^2\right).$
Noting that
\begin{eqnarray}
A_6^{I}&&= \int_0^1 \prod\limits_{i=2}^{4}d\rho _{i} \ \left(\rho _{3}\right)^{-\frac 12}\left(\rho _{4}\right)^{-\frac 12}
\ \left(1-\rho _{2}\right)^{-\frac 12}\left(1-\rho _{4}\right)^{-\frac 12}\nonumber \\
&&\ \ \ \times \ \left(1-\rho _{2}\rho _{3}\right)^{-\frac 12}\left( 1-\rho _{3}\rho _{4}\right)^
{-\frac 12}\ \left( 1-\rho _{2}\rho _{3}\rho _{4}\right)^
{-\frac 12},
\label{50}
\end{eqnarray}
and using Mellin Barnes representations, the relation $A_6^{V} =  4  A_6^I$ implies
\begin{equation}
A_6^{V}= G^{2,4}_{4,4} \left(-1\mid \matrix{\frac 12&\frac 12&\frac 12&\frac 12\cr 0&0&0&0\cr}
\right) \equiv 4\ {\sqrt \pi}\ G^{2,4}_{4,4} \left(-1\mid \matrix{\frac 12&\frac 12&\frac 12&\frac 12\cr 0&\frac 12&0&0\cr}\right)= 4 A_6^I.
\label{52}
\end{equation}

\eject

{\bf Acknowledgements}

\noindent We acknowledge discussions with A.Kaidalov, G.Korchemsky and Ch.Royon and thank S.Munier for his help. AB thanks J.Zinn-Justin for the kind hospitality at Service de Physique Th\' eorique de Saclay. This work was supported in part by the KBN Grant No 2 P03B083 08 and by PECO grant from the EEC Programme "Human Capital Mobility",
Network "Physics at High Energy Colliders", Contract No ERBICIPDCT 940613.

\eject
{\bf FIGURE CAPTION}

\vspace{1cm}

{\bf Figure 1}

{\it The 5 topologically unequivalent open-string amplitudes $A_6({\cal O})$}

The dashed lines indicate the missing exponents $K_{ij}\equiv 0,$ see text.

\input epsf
\vsize=30.truecm
\hsize=13.truecm
\epsfxsize=13.cm{\centerline{\epsfbox{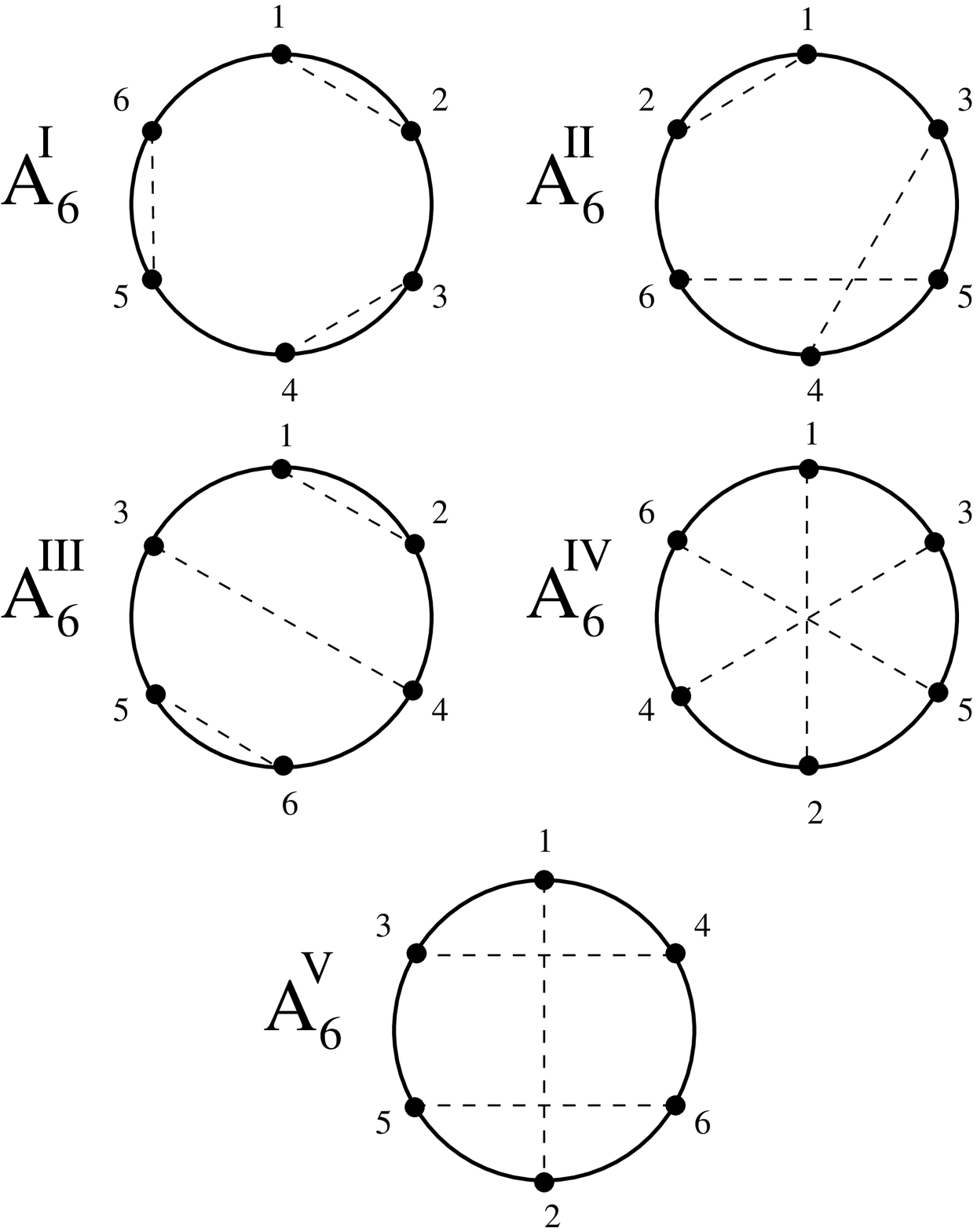}}}
\eject


\begin{thebibliography} {99}
\bibitem{na1}
The expression of the integral giving  $g_{3\cal P}$  is from \cite {pe1} and, up to a change of variables,
equivalent to the one given in \cite{bi1}.
\end{thebibliography}
\end{document}